\documentclass[1p]{elsarticle}

\pdfpagewidth=\paperwidth
\pdfpageheight=\paperheight

\usepackage{amsmath,graphicx}
\usepackage{verbatim}
\usepackage{color}
\usepackage{amsfonts,amsbsy,amssymb}
\usepackage{natmove}

\newtheorem{lemma}{Lemma}
\newdefinition{remark}{Remark}
\newproof{proof}{Proof}

\begin{document}

\author{Paul N. Patrone}
\ead{paul.patrone@nist.gov}
\author{Anthony Kearsley}
\address{National Institute of Standards and Technology}

\date{\today}

\title{Minimizing Uncertainty in Prevalence Estimates}

\begin{abstract}
Estimating prevalence, the fraction of a population with a certain medical condition, is fundamental to epidemiology.  Traditional methods rely on classification of test samples taken at random from a population.  Such approaches to estimating prevalence are biased and have uncontrolled uncertainty.  Here, we construct a new, unbiased, minimum variance estimator for prevalence.  Recent result show that prevalence can be estimated from counting arguments that compare the fraction of samples in an arbitrary subset of the measurement space to what is expected from conditional probability models of the diagnostic test.  The variance of this estimator depends on both the choice of subset and the fraction of samples falling in it.  We employ a bathtub principle to recast variance minimization as a one-dimensional optimization problem.  Using symmetry properties, we show that the resulting objective function is well-behaved and can be numerically minimized.   
\end{abstract}

\maketitle

\section{Introduction}

Estimating prevalence -- the proportion of a population that has been infected by a disease -- is a fundamental problem in epidemiology.  Nonetheless, many core mathematical issues associated with this task have only been recently discovered and understood \cite{Patrone21,Chou21}.  For example, it has long been assumed that {\it classification} of samples as positive or negative is necessary to compute the prevalence.  In Ref.\ \cite{Patrone21} we demonstrated that this is false: unbiased estimators of prevalence can be constructed from conditional probability arguments having nothing to do with classification.

The core idea of Ref.\ \cite{Patrone21} was to recognize that the probability $Q(r)$ of a diagnostic measurement outcome $r$ in a space $\Omega \subset \mathbb R^N$ is given by the convex combination
\begin{align}
Q(r) = qP(r) + (1-q)N(r), \label{eq:model}
\end{align}
where $q$ is the prevalence and $P(r)$, $N(r)$ are conditional probability density functions (PDFs) for  positive and negative populations.  Importantly, $P(r)$ and $N(r)$ can be constructed from training data and are thus known.   Taking $D$ to be an arbitrary subset of $\Omega$, one can define $q$ via
\begin{align}
q = \frac{Q_D - N_D}{P_D - N_D}, \label{eq:prevexact}
\end{align}
where $S_D$ indicates the measure of $D$ with respect to the arbitrary distribution $S$.\footnote{It is  necessary to assume that: (i) $D$ has neither zero measure nor unit measure with respect to $Q$; and (ii) the measures with respect to $P$ and $N$ are not equal.  These conditions are trivial to ensure in practice.}  Given $M$ samples  drawn at random from a population and denoted $r_j$, Eq.\ \eqref{eq:prevexact} implies that $q$ can be estimated by 
\begin{align}
q \approx \tilde q = \frac{\tilde Q_D - N_D}{P_D - N_D}, \hspace{5mm} \tilde Q_D = \frac{1}{M}\sum_{j=1}^M\mathbb I(r_j \in D),\label{eq:prevest}
\end{align}
where $\mathbb I$ is the indicator function.  
This estimate is unbiased and converges in mean square \cite{Patrone21}.

To solve the related and important problem of {\it optimal} prevalence estimation,  we minimize the variance of $\tilde q$, which is proportional to
\begin{align}
\sigma^2 = \frac{Q_D (1-Q_D)}{(P_D - N_D)^2}. \label{eq:variance}
\end{align}
Equation \eqref{eq:variance} \textcolor{black}{arises from the variance of the binomial random variable $\tilde Q_D$} but is unusual in that the denominator depends on a set that can be deformed arbitrarily.  Thus, minimization is performed with respect to both the parameter $Q_D$ as well as the $D$.  We reduce Eq.\ \eqref{eq:variance} to a one-dimensional (1D) problem by maximizing the denominator for fixed $Q_D$.  We construct this maximum in terms of a bathtub principle.

\section{Bathtub Principle}

Distinct sets $D$ may yield the same $Q_D$ but different realizations of $P_D-N_D$.  This motivates treating $Q_D$ as an independent variable, denoted by $\hat Q$ to avoid confusion.  By rewriting Eq.\ \eqref{eq:prevexact} in terms of $\hat Q$, we find a constraint 
\begin{align}
\hat Q = q(P_D - N_D) + N_D \label{eq:constraint}
\end{align}
that defines the admissible $D$ corresponding to each $\hat Q$.  Clearly the collection of sets $\{ D^\star \}_{\hat Q}$ whose elements maximize $|P_D-N_D|$ ($|\cdot|$ denotes absolute value) for a fixed $\hat Q$ is an equivalence class.  Since, each element in this class yields the same $F(\hat Q)=(P_{D^\star} - N_{D^\star})^2$ as a function of $\hat Q$, minimizing Eq.\ \eqref{eq:variance} is equivalent to solving the 1D problem
\begin{align}
\hat Q^\star = \mathop{\rm argmin}_{\hat Q} \frac{\hat Q(1-\hat Q)}{F(\hat Q)}. \label{eq:recast}
\end{align}
We next construct $F(\hat Q)$ by finding the equivalence class $\{ D^\star \}_{\hat Q}$.

Equation \eqref{eq:constraint} indicates that ``loading'' points $r$ into $D$ to increase $|P_D - N_D|$ simultaneously increases $N_D$.  We therefore wish to construct $D$ so as to increase $|P_D - N_D|$ as much as possible and $N_D$ as little as possible.  This motivates us to define the sets
\begin{align}
D_\pm &= \{r: \pm q[P(r) - N(r)] > \pm \Delta_\pm N(r) \} \cup S_\pm \label{eq:dplus} \\
\hat Q &= \int_{D_\pm} Q(r) dr. \label{eq:qplus}
\end{align}
where 
\begin{align}
S_\pm \subset \{r: q[P(r) - N(r)] = \Delta_\pm N(r) \} \label{eq:sdef}
\end{align}
and Eq.\ \eqref{eq:qplus} defines $\Delta_{\pm}$ in Eq.\ \eqref{eq:dplus}.  The $S_\pm$ are any sets satisfying both Eqs.\ \eqref{eq:sdef} and the constraints given by Eq.\ \eqref{eq:qplus}.  An optimality proof for these sets is closely related to the bathtub principle \cite{bathtub}.  For completeness and to facilitate comparison to classification methods \cite{Patrone21}, we present a proof adapted to the problem at hand.

\begin{lemma}
Let $0 < q < 1$, and assume every point $r$ has zero measure with respect to $P(r)$ and $N(r)$. For fixed $\hat Q$ satisfying $0 < \hat Q < 1$, either $D_+$ or $D_-$ maximizes $F(\hat Q)$.
\end{lemma}
{\bf Proof:} First we show that Eq.\ \eqref{eq:qplus} defines $\Delta_\pm$.  By construction, $\Delta_\pm$ satisfies the inequality $-q \le \Delta_\pm < \infty$.  Restrict attention to $\Delta_+$. Let $D(\Delta) = \{r:q[P(r) - N(r)] > \Delta N(r)\}$, where $-\infty < \Delta < \infty$.  Define $\displaystyle D(\infty)= \lim_{\Delta \to \infty}D(\Delta) = \{r: P(r) > 0, N(r)=0\}$.  Clearly,
\begin{align}
\mathcal Q(\Delta) = \left [\int_{D(\Delta)}Q(r) dr \right ]   - \hat Q 
\end{align}
is a monotone decreasing function of $\Delta$.  \textcolor{black}{Assume that $\mathcal Q(\Delta)$ crosses zero as $\Delta \to \infty$.  Then} $\mathcal Q(\Delta)$ is either continuous or discontinuous at this crossing, which defines $\Delta_+$ as the corresponding right-limit; that is $\Delta_+ = {\rm inf}_\Delta \{\Delta: \mathcal Q(\Delta) < 0\}$.  The $S_+$ can then be chosen as any subset of $\{q[P(r) - N(r)] = \Delta_+ N(r)\}$ for which $\mathcal Q(\Delta_+) + \int_{S+} Q(r)dr = 0$.  If instead $\mathcal Q(\Delta)$ does not vanish for any finite $\Delta$, then $\Delta$ is not finite and $S_+$ is a subset of $D(\infty)$. A similar argument yields existence of $\Delta_-$ and $S_-$.

Let $D$ be any set that differs from $D_+$ by positive measure not on $S_+$.  Requiring that Eq.\ \eqref{eq:constraint} also hold for $D$ yields
\begin{align}
P_{D_+} - N_{D_+} - P_D + N_D =\frac{1}{q} [N_{D/D_+} - N_{D_+/D}].
\end{align}
where $/$ is the set difference operator.  Combining the definition of $\Delta_+$ with Eq.\ \eqref{eq:constraint}, one finds that $N_{D/D_+} > N_{D_+/D}$, implying $P_{D_+} - N_{D_+} > P_D - N_D$.  A similar argument can be used to show that $D_-$ maximizes the difference $N_D - P_D$.  These differences are unique, as can be verified by the definition of $S_{\pm}$. Finally, note that $F(\hat Q)$ is maximized by $\max_{\hat Q}[P_{D_+} - N_{D_+},N_{D_-}-P_{D_-}]$. \qed

\medskip

The denominator of Eq.\ \eqref{eq:variance} can be parameterized by $\hat Q$.  In particular, we have shown that for a fixed $\hat Q$ satisfying $0 < \hat Q < 1$, one of the variances
\begin{align}
\sigma^2_+(\hat Q) = \frac{\hat Q(1-\hat Q)}{(P_{D_+}-N_{D_+})^2}, \hspace{5mm} \sigma^2_-(\hat Q) = \frac{\hat Q(1-\hat Q)}{(P_{D_-}-N_{D_-})^2} \label{eq:semioptvar}
\end{align}
minimizes $\sigma^2$.  However, we have yet to uniquely define the objective function in Eq.\ \eqref{eq:recast}, since it not clear when to use $\sigma^2_+$ or $\sigma^2_-$.  We now prove that both are equivalent.

\begin{lemma} The variances $\sigma^2_+$ and $\sigma^2_-$ satisfy the symmetry $\sigma^2_+(\hat Q) = \sigma^2_-(1-\hat Q)$.  In particular, $\sigma^2_+(\hat Q^\star)$ minimizes Eq.\ \eqref{eq:variance} if and only if $\sigma^2_-(1-\hat Q^\star)$ does. \label{lem2}
\end{lemma}

{\bf Proof:}  The numerators of $\sigma^2_+$ and $\sigma^2_-$ are invariant to the transformation $\hat Q \to 1-\hat Q$.  Also, for any $D_+$,
\begin{align}
P_{D_+} - N_{D_+} = N_{\Omega / D_+} - P_{\Omega / D_+}. \label{eq:equiv}
\end{align} 
By Eqs.\ \eqref{eq:dplus}--\eqref{eq:sdef}, the set $\Omega / D_+$ is equal to a set $D_-$ for which $\Delta_- = \Delta_+$.  Moreover, since $\hat Q$ is the $Q$-measure of domain $D_+$, the complement $\Omega / D_+$ has $Q$-measure $1-\hat Q$.  That is, $\Omega / D_+ \in \{ D_- \}_{(1-\hat Q)}$.  In light of Eq.\ \eqref{eq:equiv}, one finds $\sigma^2_+(\hat Q) = \sigma^2_-(1-\hat Q)$.  \qed


\begin{figure}
\includegraphics[width=13.5cm]{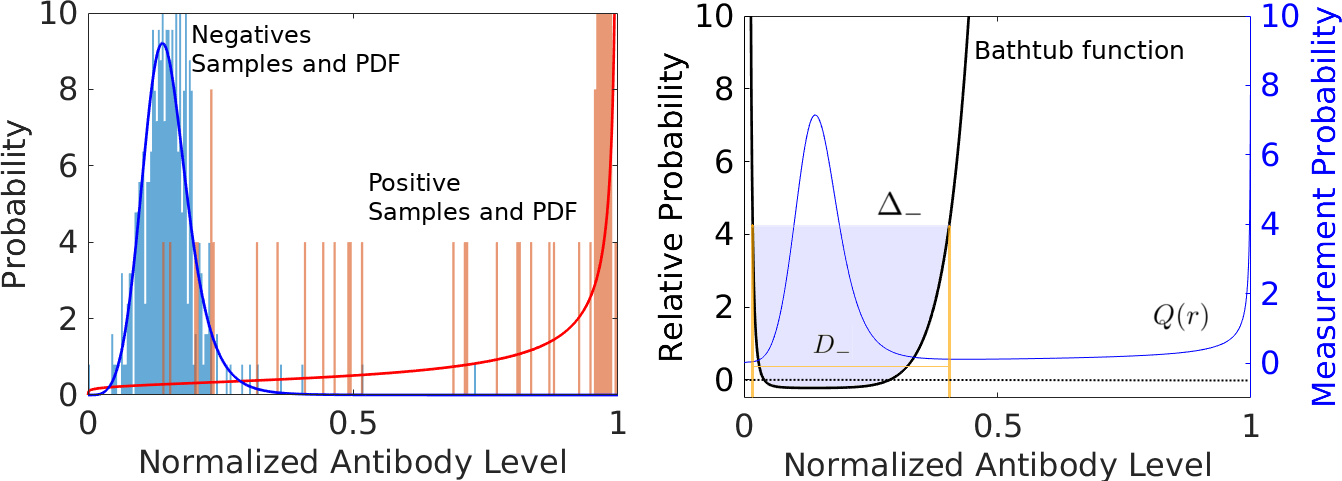}\caption{Left: data and probability models associated with a SARS-CoV-2 antibody test.  The blue and red histograms are the relative fractions of negative and positive data from an empirical study.  The continuous curves are the modeled PDFs $N(r)$ and $P(r)$. See the main text for more details.  Right: the bathtub function $q[P(r)-N(r)]/N(r)$  and $Q(r)$ for $q=30/130$.  To find the domain $D_-$, we ``fill up'' the bathtub function up to a level $\Delta_-$ for which the $Q$-measure on $D_-$ equals a specific value of $\hat Q$.}\label{fig:models}
\end{figure}

\section{Optimal Prevalence Estimation}

Lemma \ref{lem2} proves that we may treat $\sigma^2_+$ (or $\sigma^2_-$) as the objective function in Eq.\ \eqref{eq:recast}.  We now show that this objective function has desirable properties.

\begin{lemma}
Assume that: (i) any point $r$ has zero measure with respect to $P(r)$ and $N(r)$; and (ii) there is a set of positive measure with respect to $P$ for which $P(r) > N(r)$.  Then on the open domain $0 < \hat Q < 1$, the function $\sigma^2_\pm(\hat Q)$ is continuous and attains a minimum.  Moreover, $\sigma^2_{\pm} \to \infty$ as $\hat Q \to 0$ or $\hat Q \to 1$.  \label{lem3}
\end{lemma}

{\bf Proof:}  By Lemma \ref{lem2}, it is sufficient to only consider $\sigma^2_+(\hat Q)$.  Because any point $r$ has zero measure, by the definition of $D_{\pm}$, the difference $P_{D_+}-N_{D_+}$ is continuous.  Thus, $\sigma_+^2(\hat Q)$ is continuous on $(0,1)$ provided that  $P_+ - N_+ > 0$ for every $\hat Q \in (0,1)$.  To demonstrate this, assume that there exists a $\hat Q_\infty$ for which $P_{D_+} - N_{D_+} = 0$.  By assumption (ii) and the definition of $D_+$, the remaining points in $\Omega / D_+$ must have the property that $P(r) < N(r)$.  Integrating over this set shows that $P_\Omega < N_{\Omega}$, which violates the assumption that both are probability densities.  Thus $P_{D_+} - N_{D_+}$ cannot be zero in the interior of the domain, and $\sigma_+^2(\hat Q)$ is continuous on $(0,1)$.  By definition, $P_{D_+} \le \hat Q$ and $N_{D_+} \le \hat Q$, which implies that $\sigma^2_+ (\hat Q) \to \infty$ in the limit $\hat Q \to 0$.  The corresponding result in the limit $\hat Q \to 1$ is proved in the same way by considering $\hat Q \to 0$ for $\sigma^2_-$ and then using Lemma \ref{lem2}.
  
Existence of the minimum follows from divergence of $\sigma_+^2(\hat Q)$ at the boundaries and continuity in the interior $(0,1)$.  Specifically, for any $0 < \epsilon < 1/2$, $\sigma_+^2$ is continuous on $D_\epsilon=[\epsilon,1-\epsilon]$.  By the extreme value theorem, there exists a value $\hat Q_\epsilon \in D_\epsilon$ for which $\sigma_+^2(\hat Q)$ attains a minimum.  Clearly there exists an $\epsilon_0$ such that for all $\epsilon < \epsilon_0$, $\sigma_+^2(\hat Q_\epsilon)$ is constant, which defines the minimum.    \qed

{\bf Remark:}  Assumption (ii) implies that there is a set of positive measure with respect to $N$ for which $N(r) > P(r)$.  

{\bf Remark:} The assumption that there exists a set of positive measure for which $P(r) > N(r)$ is an important feature of diagnostic tests; it implies the ability to distinguish populations.  Failure to satisfy this condition is the hallmark of a useless diagnostic.

\medskip


\section{Validation and Discussion}

\subsection{Example Applied to a SARS-CoV-2 Antibody Test}

The left plot of Fig. \ref{fig:models} shows training data and probability models for a SARS-CoV-2 immunoglobulin G (IgG) receptor binding domain (RBD) assay developed in Ref.\ \cite{Gold}.  The data was normalized according to the procedure in Ref.\ \cite{Patrone21}.  Following that, we added $\varepsilon=0.01$ to all values and normalized all data by the largest positive value.  While not necessary, adding $\varepsilon$ facilitates model construction and otherwise has a negligible effect on the data, as it represents a perturbation of less than 1 \% relative to the maximum scale of the data.  We model $N(r)$ with a Burr distribution \cite{Burr} and approximate $P(r)$ as a Beta distribution.  Maximum  likelihood estimation was used to determine model parameters.  The Burr distribution was truncated to the domain $0 < r \le 1$ and renormalized to have unit probability.  The right plot of Fig.\ \ref{fig:models} shows the ``bathtub function'' $q[P(r)-N(r)]/N(r)$ for a prevalence of $q=30/130$.  The figure illustrates that for a given $\Delta_-$, the domain $D_-$ corresponds to the set of all $r$ for which $q[P(r)-N(r)]/N(r)<\Delta_-$.  The corresponding value of $\hat Q$ is given by the integral of $Q(r)$ over $D_-$.

\begin{figure}
\includegraphics[width=13.5cm]{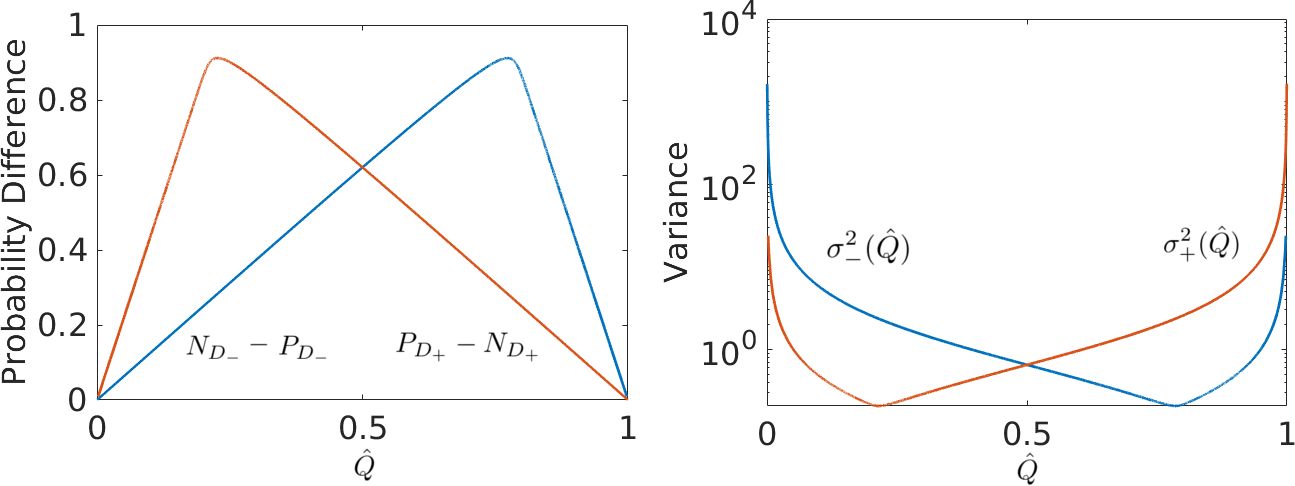}\caption{Left: The differences $P_{D_+}-N_{D_+}$ and $N_{D_-}-P_{D_-}$ as a function of $\hat Q$.  Note that the differences are positive on the interior of the domain and reflect the symmetry implied by Eq.\ \eqref{eq:equiv}.  Right: The variances $\sigma^2_+(\hat Q)$ and $\sigma^2_-(\hat Q)$ versus $\hat Q$.  Note the divergence of both functions as $\hat Q \to 0$ and $\hat Q \to 1$, consistent with Lemma \ref{lem3}.  Moreover, the variances exhibit the same symmetry as in the left plot.}\label{fig:pdiff}
\end{figure}

Figure \ref{fig:pdiff} illustrates various results derived in Lemma \ref{lem3}.  The left plot embodies the symmetry argument expressed in Eq.\ \eqref{eq:equiv}, as well as the fact that the difference $P_{D_+}-N_{D_+}$ and $N_{D_-}-P_{D_-}$ are positive on the open set $0 < \hat Q < 1$.  The right plot illustrates that $\sigma^2_+(\hat Q) = \sigma^2_-(1-\hat Q)$, as well as the divergence when $\hat Q \to 0$ and $\hat Q \to 1$.  Moreover, the objective function is continuous and bounded from below.

\subsection{Limitations and Open Directions}

The analysis presented herein does not refer to positive or negative populations.  The functions $P(r)$ and $N(r)$ could have described two different populations having nothing to do with diagnostics.  Our main result has implications for broader problems associated with estimating relative fractions of different populations.  

Our method suffers a need to empirically model training data, which may introduce uncertainty associated with the choices of distributions used.  Addressing such tasks is necessary to fully understand all sources of uncertainty in prevalence estimates.  The analysis herein is idealized insofar as it assumes that $P(r)$ and $N(r)$ are known exactly.  

In a related vein, construction of the optimal prevalence estimation domains requires a priori knowledge of $q$ itself.  Since this quantity is often what is being estimated, a practical algorithm for minimizing uncertainty in $q$ would be to guess a trial domain based on any prior information, estimate $q$, and update the domain.  While this approach will likely never converge in a mathematical sense, it should provide reasonable estimates motivated by the theory herein.

{\it Acknowledgements:} This work is a contribution of the National Institutes of Standards and Technology and is therefore not subject to copyright in the United States.

{\it Use of all data deriving from human subjects was approved by the NIST Research Protections Office.}

\bibliographystyle{elsarticle-num}
\bibliography{Prev_AML}

\end{document}